\documentclass[reprint,amsmath, superscriptaddress, amssymb, aps, prl]{revtex4-2}
\usepackage{lipsum}  
\usepackage{stackengine}
\usepackage{graphicx}
\usepackage{hyperref}
\usepackage{amsmath}
\usepackage{soul}
\usepackage{dcolumn}
\usepackage{lipsum}  
\usepackage{float}
\usepackage{comment}
\newcommand\thefontsize{The current font size is: \f@size pt}
\usepackage{bm}
\usepackage{color}
\usepackage[normalem]{ulem}
\usepackage{lineno}
\begin{document}
\renewcommand{\thesection}{\Roman{section}}
\preprint{APS/123-QED}
\title{Impact of an electron Wigner crystal on exciton propagation}
\author{Daniel Erkensten}
\affiliation{Department of Physics, Philipps-Universit{\"a}t Marburg, 35037 Marburg, Germany}
\affiliation{mar.quest|Marburg
Center for Quantum Materials and Sustainable Technologies,
35032 Marburg, Germany}
\author{Alexey Chernikov}
\affiliation{Institute of Applied Physics and Würzburg-Dresden Cluster of Excellence ct.qmat,
TU Dresden, 01187 Dresden, Germany}
\author{Ermin Malic}
\affiliation{Department of Physics, Philipps-Universit{\"a}t Marburg, 35037 Marburg, Germany}
\affiliation{mar.quest|Marburg
Center for Quantum Materials and Sustainable Technologies,
35032 Marburg, Germany}
\begin{abstract}
The strong Coulomb interaction in 2D materials facilitates the formation of tightly bound excitons and charge-ordered phases of matter. A prominent example is the formation of a crystalline phase from free charges due to mutual Coulomb repulsion, known as the Wigner crystal. While exciton-electron interactions have been used as a sensor for Wigner crystallization, its impact on exciton properties has been poorly understood so far. Here, we show that the weak potential induced by periodically ordered Wigner crystal electrons has a major impact on exciton propagation, albeit having only a minor influence on exciton energy. The effect is tunable with carrier density determining the  Wigner crystal confinement and temperature via thermal occupation of higher subbands. Our work provides microscopic insights into the interplay between excitons and charge-ordered states identifying key signatures in exciton transport, and establishes a theoretical framework for understanding exciton propagation in the presence of strong electronic correlations.
\end{abstract}
\maketitle
\begin{figure}[t!]
    \centering
    \includegraphics[width=\linewidth]{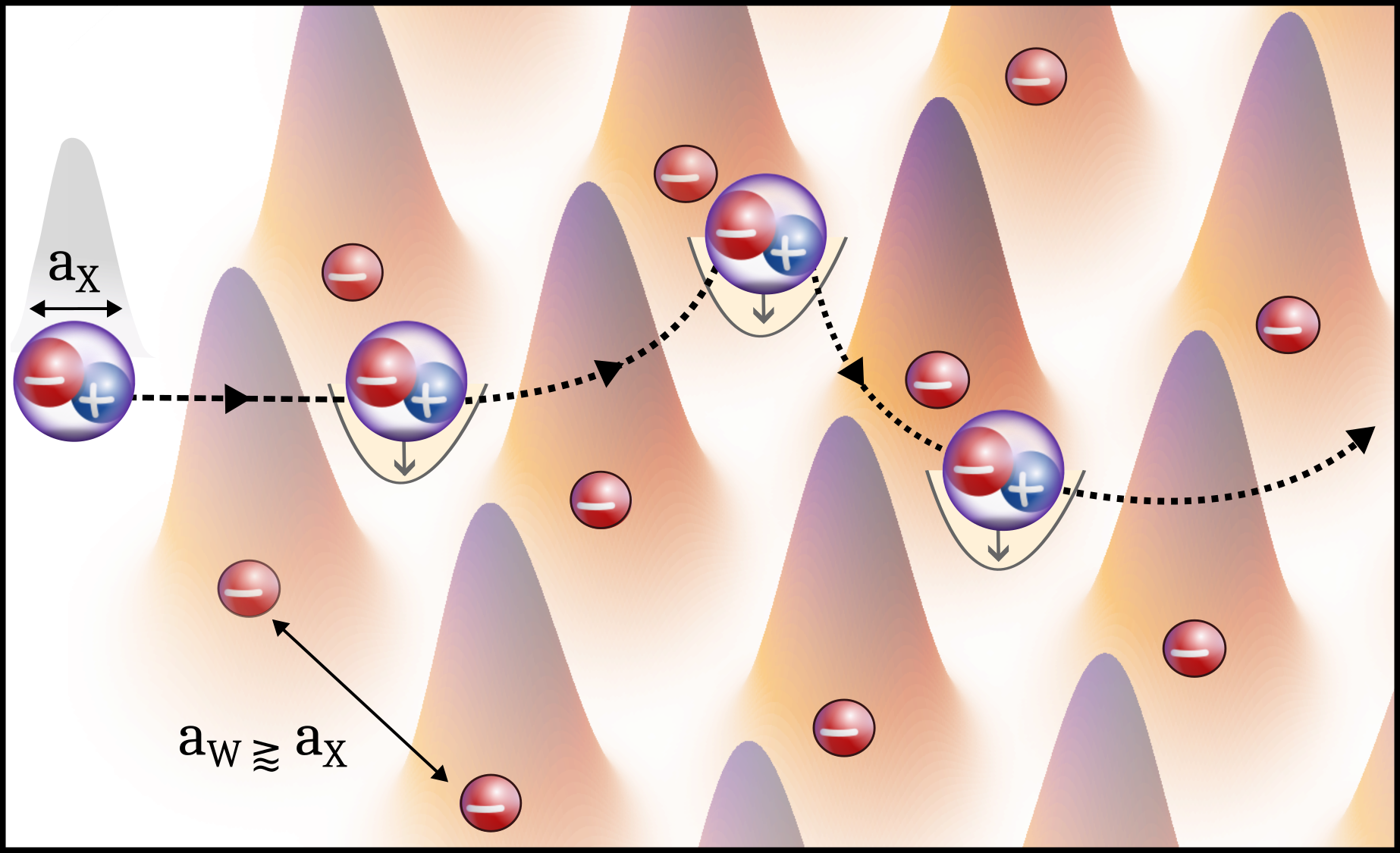} 
    \caption{Schematic illustration of exciton transport in the presence of an electron Wigner crystal. The periodic  arrangement of electrons in a Wigner crystal lattice gives rise to a potential that excitons can be trapped in, slowing down their propagation. The colored Gaussian curves illustrate the real-space charge density of  electrons confined in a Wigner lattice with the periodicity set by the Wigner lattice constant $a_W$ that is generally comparable to the spatial extent of the exciton $a_X$.}
    \label{schematicfig}
\end{figure}

\emph{Introduction.---}
In recent years, transition-metal dichalcogenides (TMDs) have been shown to offer an exceptional material platform to study rich exciton phenomena \cite{perea2022exciton, wang2018colloquium, mueller2018exciton, siday2022ultrafast} and strongly correlated states of matter \cite{smolenski2021signatures, regan2020mott, zhou2021bilayer, tang2022dielectric, cai2023signatures}. The remarkably strong Coulomb interaction in these materials facilitates the formation of tightly bound excitons that are stable even at room temperature, as well as higher-order charge complexes, such as trions \cite{mak2013tightly, ross2013electrical, perea2024trion,courtade2017charged}, biexcitons \cite{steinhoff2018biexciton, katsch2020optical, brem2024optical}, Wigner crystals  \cite{smolenski2021signatures, brem2022terahertz, sung2025electronic}, or Mott insulating states in TMD-based heterostructures \cite{regan2020mott, zhou2021bilayer, li2021imaging, tang2022dielectric}. The century-old prediction that an electron gas forms a triangular crystal at sufficiently low carrier densities and temperatures \cite{PhysRev.46.1002}, resulting in a Wigner crystal (Fig. \ref{schematicfig}), has recently been verified by optical spectroscopy on a two-dimensional solid-state platform in MoSe$_2$ monolayers \cite{smolenski2021signatures, sung2025electronic, chen2025terahertzelectrodynamicszerofieldwigner}. 

Despite these advances, little attention has been devoted to understanding the complex interplay between optically excited electron-hole pairs and electrons confined to a Wigner crystal lattice in atomically thin semiconductors in the context of exciton propagation and dynamics. Only recently, exciton propagation experiments in the vicinity of correlated states were reported using externally defined periodic potentials in moir{\'e} heterostructures, where the charge trapping is determined by the spatial variation of atomic registries \cite{yan2024anomalously, upadhyay2024giant, deng2025frozen}. In particular, exciton propagation has been shown to be suppressed in the presence of generalized Wigner crystal states due to effective exciton-electron scattering \cite{yan2024anomalously}, whereas propagation in the vicinity of Mott insulating states has been shown to be complex and could be enhanced due to strong exciton-electron repulsion in mixed exciton-electron lattices \cite{upadhyay2024giant}. The long-range dipolar repulsion between moir{\'e} excitons could also freeze the propagation of excitonic Mott insulating states \cite{deng2025frozen}. In the absence of external potentials, exciton diffusion has been recently studied in the presence of a Fermi sea of \emph{free} carriers demonstrating a non-monotonous dependence with respect to the carrier density. This was explained by considering two distinct regimes governed by elastic exciton-electron scattering and trion formation, respectively \cite{wagner2023diffusion}. However, the impact of a spontaneous symmetry breaking in the distribution of free charges forming a Wigner crystal on exciton transport has remained a major open question.

In this work, we introduce a bridge between the physics of electronic correlations and the field of exciton transport. In particular, we investigate exciton diffusion in the presence of an electronic Wigner crystal based on a microscopic and material-specific  theory that could be generalized to any type of charge-ordered states. We consider the exemplary case of a MoSe$_2$ monolayer, which offers an ideal platform that hosts low-energy bright excitons and where the formation of electron Wigner crystal states has been experimentally verified \cite{smolenski2021signatures}.  Despite the fact that the exciton-electron interaction is weak leading to only tiny exciton energy shifts in the sub-meV range \cite{smolenski2021signatures, shimazaki2021optical}, we predict a significant flattening of exciton bands and a two-fold decrease in the exciton diffusion coefficient under experimentally realistic conditions for Wigner crystallization. As excitons become subject to a periodic potential induced by the periodically arranged Wigner electrons (Fig. 1), they feel the shallow potential from Wigner electrons slowing down their propagation - in analogy to exciton trapping in moir{\'e} potential pockets in twisted or lattice-mismatched TMD heterostructures \cite{tran2019evidence, huang2022excitons, yuan2020twist, brem2020tunable}. Here, however, the strong impact of the correlated states on exciton properties is a consequence of the Coulomb interaction alone. Interestingly, we find that exciton diffusion becomes more efficient at higher carrier densities at cryogenic temperatures - in direct contrast to what is expected from excitons scattering with free electrons as shown in recent transport experiments on doped monolayer TMDs \cite{wagner2023diffusion}. Altogether, our work demonstrates substantial impact of Wigner crystal formation on exciton transport that should be accessible at realistic experimental conditions. \\ 

\emph{Microscopic model.---}
 To microscopically model exciton propagation in the presence of an electronic Wigner crystal, we start from the many-particle exciton Hamilton operator 
\begin{equation}
H=\sum_{\mathbf{Q}}E_{\mathbf{Q}}X^{\dagger}_{\mathbf{Q}}X_{\mathbf{Q}}+\sum_{\mathbf{q}, \mathbf{Q}}V_{x-e}(\mathbf{q})\rho_{e}(\mathbf{q})X^{\dagger}_{\mathbf{Q}+\mathbf{q}}X_{\mathbf{Q}} \ , 
\label{hamiltonian}
\end{equation}
where the first term contains the free parabolic center-of-mass exciton dispersion $E_{\mathbf{Q}}=\frac{\hbar^2|\mathbf{Q}|^2}{2M}$ with $\mathbf{Q}$ being the center-of-mass momentum and $M=m^{*}_e+m^{*}_h$  the total exciton mass with $m^{*}_e$ and $m^{*}_h$ as electron and hole masses, respectively. The second term is of key importance and describes the mean-field interaction of excitons with Wigner crystal electrons, where the exciton-electron interaction $V_{x-e}(\mathbf{q})$ is weighted by the Wigner electron momentum charge density $\rho_e(\mathbf{q})$. The exciton-electron interaction is approximated by a contact-like interaction in real space \cite{efimkin2021electron, sidler2017fermi, efimkin2018exciton}, such that $V_{x-e}(\mathbf{q})=v_{x-e}$ with $v_{x-e}$ being the long-wavelength limit of the fully microscopic and material-specific exciton-electron interaction \cite{efimkin2021electron} (see Supplemental Material (SM) Section I for details). The Wigner electron momentum density $\rho_e$ is assumed to have a Gaussian density profile in momentum and real space, that is, $\rho_e(\mathbf{r}) =\frac{1}{2\pi\xi^2}\sum_{n}\mathrm{e}^{-|\mathbf{r}-\mathbf{R}_n|^2/\xi^2}$ with $\mathbf{R}_n$ being real-space Wigner lattice vectors and $\xi$ describing the spatial extension of the electron wave function around the lattice sites. The latter is defined as a fraction of the Wigner lattice period (set by the carrier density) and is obtained from a variational approach by minimizing the Hartree Coulomb repulsion energy and the kinetic energy of carriers \cite{joy2022wigner} (see SM Section II). 

The many-particle Hamilton operator in Eq. \eqref{hamiltonian} describes excitons in
the presence of an effective periodic potential induced by Wigner crystal electrons (Fig. \ref{schematicfig}). As such, the Wigner crystal potential is expected to renormalize the free parabolic exciton band structure. It can result in a density-dependent flattening of exciton bands even in the absence of static external lattice-induced potentials \cite{brem2023bosonic}. In particular, the energy spectrum of the Hamiltonian is now given by a series of subbands $\eta$ defined in the mini-Brillouin zone (mBZ) spanned by the reciprocal Wigner lattice vectors (SM Section III). 

\begin{figure}[t!]
    \centering
    \includegraphics[width=\linewidth]{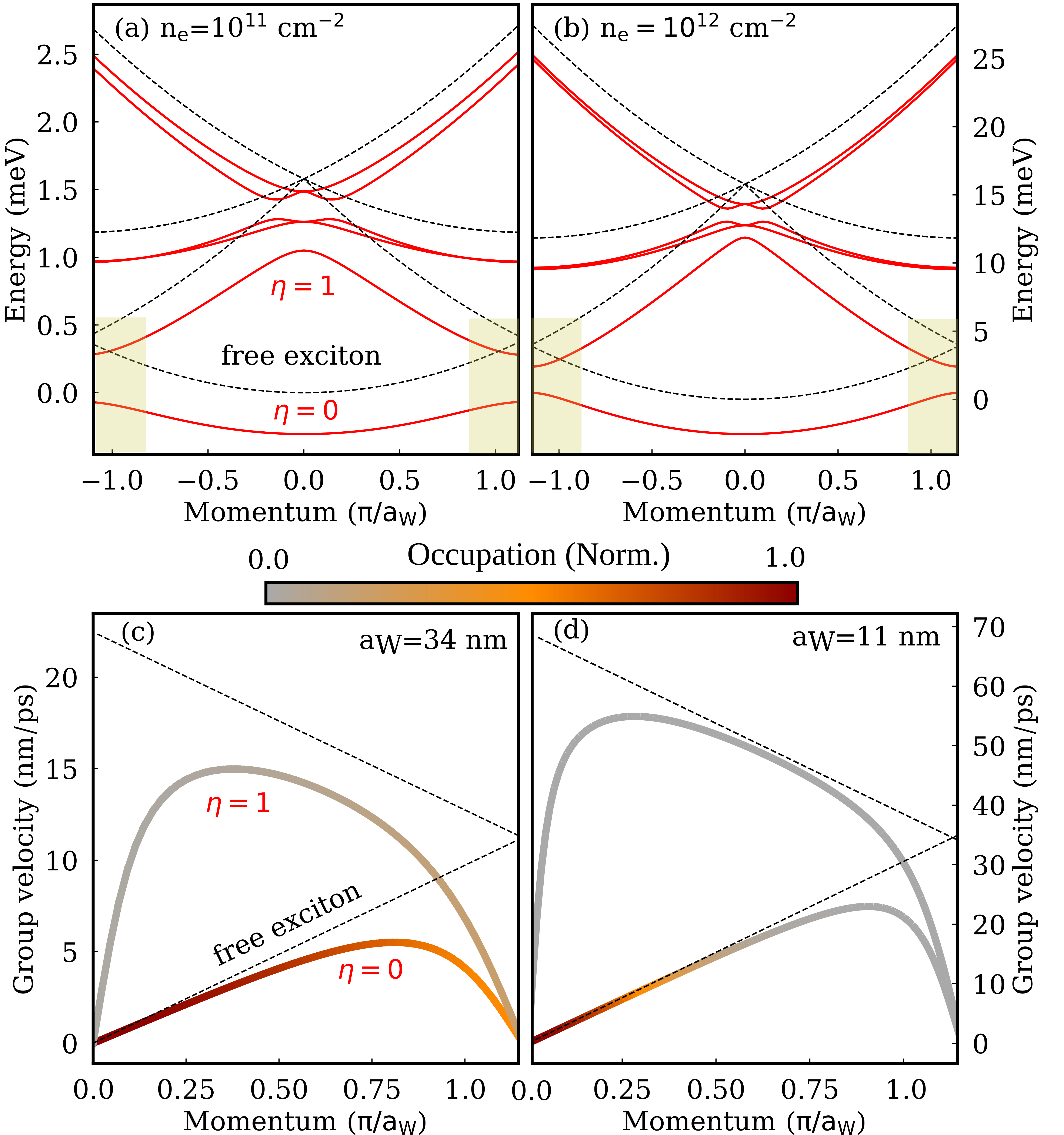}
    \caption{Exciton band structures and group velocities along a horizontal cut of the Wigner crystal Brillouin zone in hBN-encapsulated MoSe$_2$ monolayers for (a) low carrier density  $n_e=10^{11}$ $\mathrm{cm}^{-2}$ (corresponding to a Wigner lattice period $a_W\approx{34}$ nm) and (b) high carrier density $n_e=10^{12}$ $\mathrm{cm}^{-2}$ ($a_W\approx{11}$ nm). The bare bands without a Wigner crystal potential are shown with dashed lines. The flattening of the bands at the edges of the Brillouin zone (yellow areas in [(a)-(b)]), results in quenched group velocities at larger momenta [(c)-(d)]. The group velocities are superimposed by the corresponding exciton occupations for the two lowest-lying bands.}
    \label{bandstructure}
\end{figure}
We diagonalize the Hamiltonian in Eq. \eqref{hamiltonian} for the exemplary case of an hBN-encapsulated MoSe$_2$ monolayer. The resulting renormalized exciton band structure is displayed in Fig. \ref{bandstructure} along a horizontal cut of the Wigner mini-Brillouin zone and shown for the two different electron densities of $n_e=10^{11}$ $\mathrm{cm}^{-2}$ [(a)] and $n_e=10^{12}$ $\mathrm{cm}^{-2}$ [(b)], corresponding to the Fermi energies $E_F=\hbar^2 \pi n_e/m_e^*$ of 0.5 meV up to 5 meV, respectively. Importantly, we note that the higher-lying excitonic resonances  can be understood as umklapp processes, where the energy of the umklapp-scattered excitons is provided by the reciprocal Wigner lattice vector  \cite{smolenski2021signatures, shimazaki2021optical}. Concretely, we obtain a splitting $\Delta E\approx 0.5$ meV between the first umklapp-scattered exciton state $\eta=1$ and the ground state $\eta=0$  at $n_e=10^{11}$ $\mathrm{cm}^{-2}$, which grows linearly with the increasing carrier density. This can be well understood in the limit of a weak exciton-electron interaction, where the energy splitting is determined by the exciton kinetic energy at the momentum corresponding to the magnitude of a reciprocal Wigner lattice vector resulting in $\Delta E=\frac{\hbar^2 |\mathbf{G}_W|^2}{2M}\sim a_W^{-2}\sim n_e$ \cite{shimazaki2021optical, smolenski2021signatures}. Here, $\mathbf{G}_W$ is the (first-shell) reciprocal Wigner lattice vector with $|\mathbf{G}_W|\propto a^{-1}_W$, where $a_W$ is the Wigner lattice period. The obtained splitting is in good agreement with the umklapp resonance energy measured previously \cite{smolenski2021signatures}.  \\

We find a generally small energy renormalization of exciton energy in the presence of a Wigner crystal potential due to the weak exciton-electron interaction, cf. the solid red and dashed black lines in Figs. \ref{bandstructure}(a)-(b) denoting the renormalized and bare exciton bands, respectively. Interestingly, we show that exciton bands become flattened around the edges of the Brillouin zone for the lowest-lying $\eta=0$ subband (yellow areas in Fig. \ref{bandstructure}(a)-(b)). The band flattening is more pronounced at low densities, which is traced back to the stronger localization of Wigner crystal electrons at low densities, i.e., the spatial extent $\xi$ of Wigner electrons is much smaller than the Wigner period $a_W$. In particular, we find that the Lindemann ratio $\xi/a_W$ scales with $a_W^{-1/4}\sim n_e^{1/8}$ \cite{joy2022wigner}. The delocalization of Wigner electrons at elevated carrier densities also results in delocalized exciton wave functions, and thereby in more disperse bands. We show this by performing a harmonic approximation of the exciton-electron interaction potential around its minima, such that the Wannier function associated with the lowest-lying exciton subband becomes a Gaussian with the width $a_x\propto \xi$ (see SM Section IV and Fig. \ref{schematicfig}). Since $\xi/a_W\propto n_e^{1/8}$, we find that excitons are more localized at lower electron densities and we expect the lowest-lying exciton subband to become more disperse as $n_e$ increases.

The flattening of exciton bands at finite momenta results in suppressed group velocities $v^{\eta}_Q=\frac{1}{\hbar}|\nabla_{\mathbf{Q}}\epsilon^{\eta}_{\mathbf{Q}}|$ compared to the free exciton case with $v_{\mathbf{Q}}=\frac{\hbar Q}{M}$, cf. Fig. \ref{bandstructure} (c)-(d). Due to the small size of the Brillouin zone at the lower density $n_e=10^{11}$ $\mathrm{cm}^{-2}$ ($a_W\approx{34} $ nm), the flat parts of the band structure exhibiting small group velocities are strongly occupied even at cryogenic temperatures, as shown in Figs. \ref{bandstructure} (c)-(d), where a thermal Boltzmann distribution is mapped on the momentum-dependent group velocities. Since exciton diffusion depends not only on the group velocity, but also on the occupation of exciton states (cf. Eq. \eqref{diffcoeff}), it is expected that the occupied flat parts contribute significantly to exciton transport. In contrast, at higher densities, only a small part of the states in the much larger Brillouin zone is populated ($a_W\approx{11} $ nm), i.e., the flat regions at higher momenta remain unoccupied (Fig. \ref{bandstructure} (c)) and do not contribute to exciton diffusion.
As a result, considering the changes in the group velocity and subband occupation, exciton diffusion is expected to be slowed down at lower carrier densities. \\

\emph{Exciton propagation.---} To obtain microscopic access to exciton propagation in the vicinity of correlated states, we calculate the exciton diffusion coefficient $D$. The latter is derived using the Wigner function formalism \cite{hess1996maxwell} and by applying the relaxation-time approximation yielding \cite{hess1996maxwell, rosati2020negative, perea2019exciton, meneghini2025spatiotemporal}
\begin{equation}
D=\frac{1}{2}\sum_{\eta}\int_{\mathrm{mBZ}} \mathrm{d}^2\mathbf{Q}\tau^{\eta}_{\mathbf{Q}}(v^{\eta}_{\mathbf{Q}})^2 N^{\eta}_{\mathbf{Q}} \ , 
\label{diffcoeff}
\end{equation}
where $\tau^{\eta}_{\mathbf{Q}}$  describes the relaxation time due to exciton-phonon scattering. Furthermore, $v^{\eta}_{\mathbf{Q}}=\frac{1}{\hbar} |\nabla_{\mathbf{Q}}\epsilon^{\eta}_{\mathbf{Q}}|$ corresponds to the band-specific group velocity obtained directly from the renormalized exciton band structure $\epsilon^{\eta}_{\mathbf{Q}}$. Moreover, the exciton occupation  $N^{\eta}_{\mathbf{Q}}$ is estimated by a Boltzmann distribution. To derive Eq. \eqref{diffcoeff}, we perform a zone-folding of the exciton dispersion into the mini-Brillouin zone of the Wigner lattice. Importantly, considering low temperatures, such that only the lowest exciton subband is occupied, and assuming that the band is parabolic and  $\tau^{\eta}_{\mathbf{Q}}\approx{\tau}$, the diffusion coefficient reduces to the well-known semi-classical expression $D\approx \frac{k_B T \tau}{M}$ with $T$ as the temperature of the excitonic system \cite{wagner2021nonclassical}. The relaxation time $\tau^{\eta}_{\mathbf{Q}}$ is obtained from microscopically calculated exciton-phonon scattering rates explicitly taking into account the superlattice potential given by the periodic potential induced by the Wigner crystal electrons (SM Section V) \cite{meneghini2024excitonic}. We note that the semi-classical approximation of exciton transport is expected to hold at the considered cryogenic temperatures and that quantum corrections become important at higher temperatures \cite{wagner2021nonclassical, glazov2020quantum}. Furthermore, we remark on the exciton transport being diffusion-driven rather than hopping-driven due to the exciton-electron potential energy being similar to the thermal energy resulting in similar length scales for the localization of excitons and the Wigner lattice period (Fig. \ref{schematicfig} and SM Section IV). The material-specific input parameters used for transport calculations including electron and hole masses, dielectric constants, electron-phonon coupling strength, and phonon energies, are extracted from \emph{ab-initio} calculations \cite{kormanyos2015k, PhysRevB.90.045422} and provided in the SM Section VII.

\begin{figure}[t!]
    \centering
    \includegraphics[width=\linewidth]{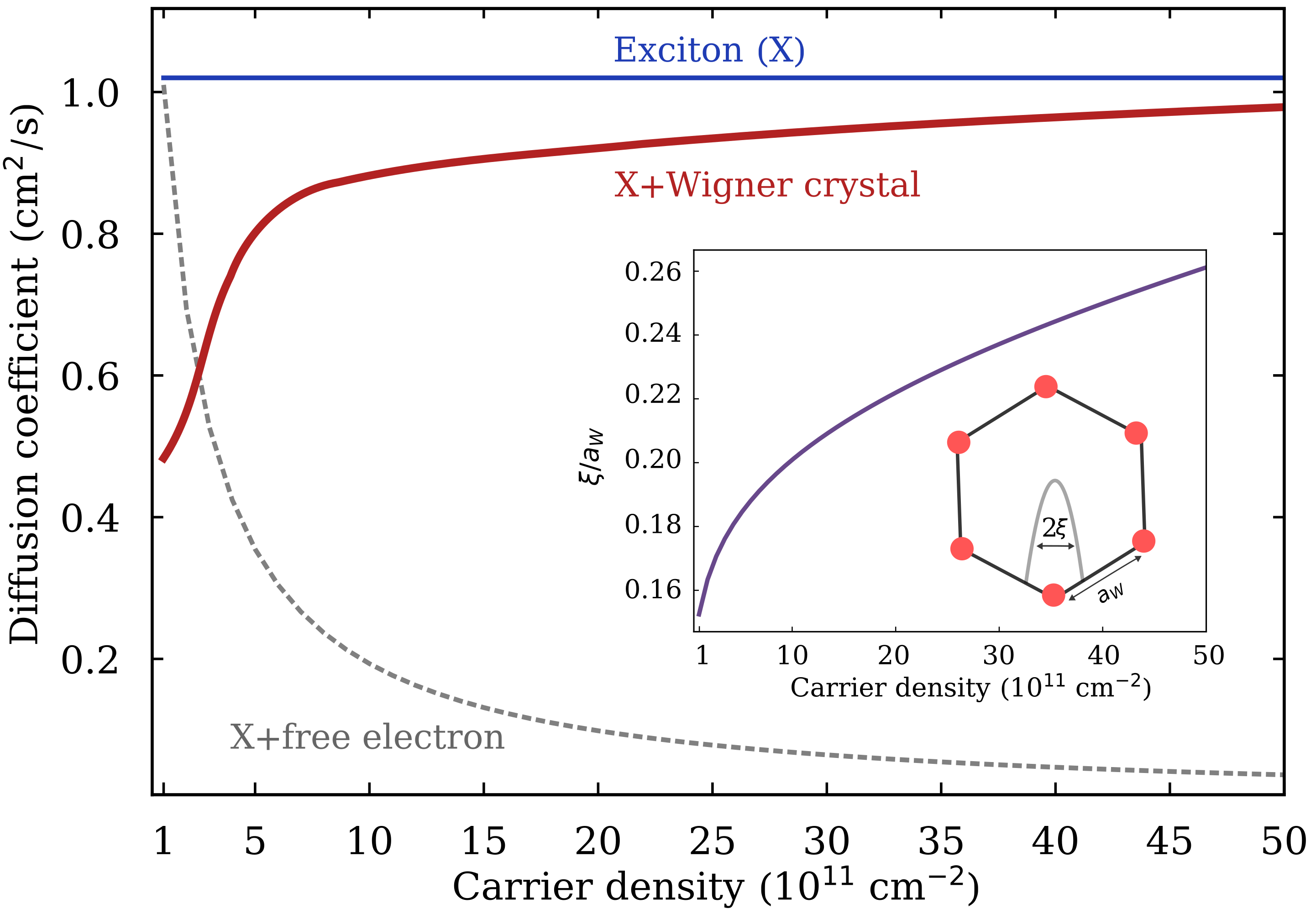}
    \caption{Density-dependent exciton diffusion coefficient in the presence of an electron Wigner crystal in hBN-encapsulated MoSe$_2$ monolayers at cryogenic temperatures ($T$=4 K). At low carrier densities, the Wigner crystal electrons are strongly localized (inset), i.e., their spatial extent $\xi$ is much smaller than the Wigner lattice period $a_W$. This gives rise to the emergence of flattened exciton bands and suppressed exciton diffusion. As the carrier density is increased, Wigner electrons become less confined, exciton bands become increasingly parabolic, and exciton diffusion approaches the limit of free excitons (blue line). In strong contrast, free electron-exciton scattering gives rise to a decrease in the exciton diffusion as a functon of carrier density (dashed gray line). }
    \label{diffusion}
\end{figure}
We now evaluate the exciton diffusion coefficient (Eq. \eqref{diffcoeff}) for the exemplary case of a bright (KK) exciton in a MoSe$_2$ monolayer. In Fig. \ref{diffusion}, the diffusion coefficient is shown as a function of Wigner electron density. Considering free exciton propagation, we find that the exciton diffusion coefficient is density independent (blue line in Fig. \ref{diffusion}) and is given by $D\approx{1}$ $\mathrm{cm}^2/\mathrm{s}$. Taking into account the impact of a Wigner crystal potential on the propagation of excitons, we reveal an intriguing drop in the exciton diffusion coefficient at low carrier densities down to $D\approx{0.5}$ $\mathrm{cm}^2/s$ at the lowest considered density of $n_e=10^{11}$ $\mathrm{cm}^{-2}$  (red line in Fig. \ref{diffusion})). This characteristic behavior has the opposite density dependence  compared to the diffusion obtained for excitons scattering with free electrons (dashed grey line in Fig. \ref{diffusion}). The latter was obtained within a Fermi-polaron approach further discussed in the SM Section VI.  

To obtain a better understanding of the predicted density dependence of the diffusion coefficient in the presence of a Wigner crystal, we come back to Eq. \eqref{diffcoeff}. Here, the diffusion coefficient is determined by the exciton group velocity, exciton occupation, and exciton-phonon scattering time. The latter is enhanced at lower densities reflecting the decrease in the number of scattering channels when exciton bands are flattened, and therefore a boost in exciton transport at lower carrier densities would be expected - in contrast to the behaviour shown in Fig. \ref{diffusion}. We find that the predominant density dependence of the diffusion coefficient can be traced back to the suppression of group velocities and the occupation of flat parts in the excitonic band structure (Fig. \ref{bandstructure}(c)). At low carrier densities, the Wigner crystal electrons are well-localized, i.e. their spatial extent $\xi$ is small compared to the Wigner lattice constant (inset in Fig. \ref{diffusion}). The degree of localization is described by the Lindemann parameter $\xi/a_W$, which is growing with increasing density: the smaller the Lindemann parameter, the more localized Wigner electrons and the flatter exciton bands (Fig. \ref{bandstructure}(a)). Upon increasing the carrier density,  Wigner crystal electrons start to overlap and become delocalized (consistent with the Wigner crystal approaching quantum melting). As a consequence, excitons become more mobile and their bands more parabolic. In addition, the size of the Brillouin zone increases with density such that the flatter parts of the band structure are no longer populated (Fig. \ref{bandstructure}(b)). This eventually leads to a diffusion coefficient that corresponds to the case of free excitons (blue line in Fig. \ref{diffusion}). 

For a comparison with experimentally realistic scenarios, it is important to note that while Wigner crystal melting is not included in the microscopic model, it is expected to take place only in the upper range of the considered density regime ($n_e\gtrsim 5\cdot 10^{11}$ $\mathrm{cm}^{-2}$) \cite{smolenski2021signatures, erkensten2024stability}. We thus expect our findings to hold and be experimentally accessible for the considered free carrier densities, predicting a drastic drop in the diffusion coefficient in the low-density limit. The predicted, characteristic density dependence of the exciton diffusion coefficient in Fig. \ref{diffusion} strongly contrasts the density dependence of exciton diffusion for the case of a Fermi sea of free electrons. Applying a Fermi-polaron approach, we indeed find a rapid decrease in exciton diffusion as a function of carrier density instead. This reflects the efficient scattering between excitons and free electrons \cite{wagner2023diffusion}. The decrease in exciton diffusion is a qualitatively different behavior compared to the predicted exciton propagation in the presence of a Wigner crystal, where we find a monotonous increase in the diffusion coefficient as a function of carrier density. Thus, the drop in the exciton diffusion at low densities (red line in Fig. \ref{diffusion}) is identified as a clear hallmark for the Wigner crystallization. 

\begin{figure}[t!]
    \centering
    \includegraphics[width=\linewidth]{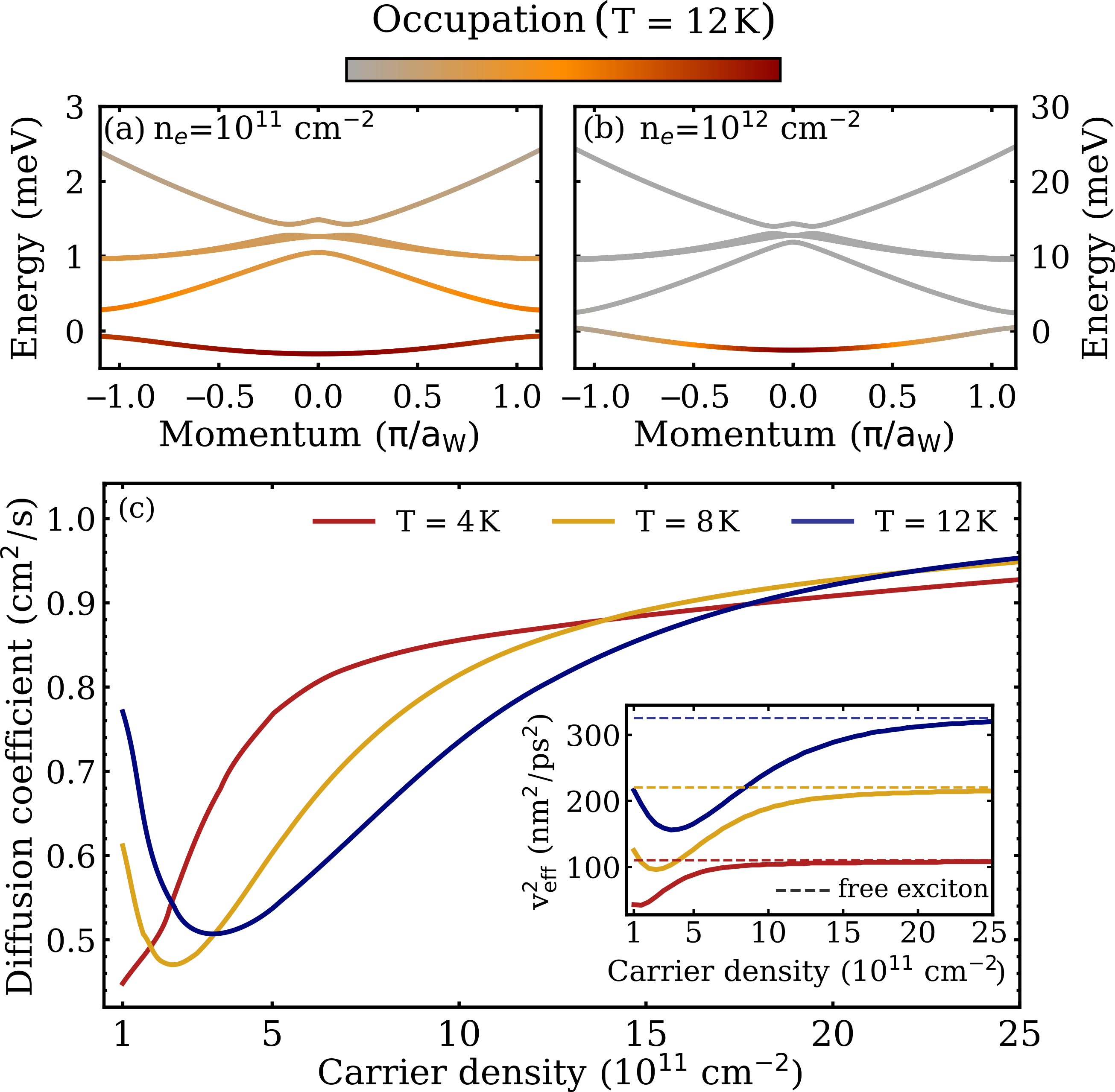}
    \caption{Temperature-dependent exciton diffusion in the presence of a Wigner crystal. Exciton band structure with the bands overlaid by the exciton occupation at $T=12$ K, revealing (a) a large occupation of higher-lying exciton bands at low densities and (b) their negligible occupation at high densities. (c) Exciton diffusion coefficient  at three  different temperatures. For $T=8$ and 12 K,  exciton diffusion becomes non-monotonous as a function of carrier density, reflecting the behaviour of the effective thermal group velocity (inset).}
    \label{tempdiffusion}
\end{figure}

 Now, we study the density dependence of the exciton diffusion coefficient at different temperatures.   At $T=4$ K, excitons predominantly occupy only the energetically lowest subband (Fig. \ref{bandstructure}(c)-(d)) even at low densities, where the separation between different subbands is very small. At elevated temperatures, also higher-lying bands ($\eta\geq 1$) become populated as shown in Fig. \ref{tempdiffusion}(a) for the case of $T=12$ K and $n_e=10^{11}$ $\mathrm{cm}^{-2}$. These bands are more disperse and have  higher group velocities. As a consequence, for a fixed low density, the diffusion coefficient is expected to be larger at elevated temperatures. Furthermore, at higher densities ($n_e=10^{12}$ $\mathrm{cm}^{-2}$), we find that only the first exciton subband is occupied at $T=12$ K (Fig. \ref{tempdiffusion} (b)), but a larger part of the Brillouin zone is populated compared to the case of $T=4$ K (Fig. \ref{bandstructure}(d)). 
 
 In Fig. \ref{tempdiffusion}(c), we show the density-dependent exciton diffusion coefficient for three different temperatures of $T=4,8$ and 12 K. Note that for  $T>12$ K, the Wigner crystal has been shown to melt \cite{smolenski2021signatures}. For higher temperatures considered here ($T=8$ K and $T=12$ K, yellow and blue lines, respectively), we again find a drop in the exciton diffusion coefficient, when the density is reduced. However, in contrast to the case of $T=4$ K discussed so far, the diffusion coefficient becomes non-monotonous and increases again for very low densities. Thus, the diffusion coefficient exhibits a minimum at a specific temperature-dependent carrier density. This can be directly traced back to density-dependent minima found in the squared effective group velocity (thermally averaged velocity), $v^2_{\mathrm{eff}}=\sum_{\eta, \mathbf{Q}}(v^{\eta}_{\mathbf{Q}})^2 N^{\eta}_{\mathbf{Q}}$, see the inset in Fig. \ref{tempdiffusion}(c). At low carrier densities, several subbands with large group velocities are occupied (Fig. \ref{tempdiffusion}(a)). As the density increases, the energy separation between the subbands becomes larger, such that only the lowest-lying band is eventually populated (Fig. \ref{tempdiffusion}(b)) leading to a decrease in the squared effective group velocity. The higher the density, the more parabolic the relevant subband becomes, leading to an increase in the squared effective group velocity. In the limit of high carrier densities, the thermally averaged squared group velocity approaches $v_{\mathrm{eff}}^2=\frac{2 k_B T}{M}$ - as expected from the equipartition theorem (dashed lines in inset in Fig. \ref{tempdiffusion}(c)). Here, the diffusion coefficient approaches the value of free excitons.  Overall, density-dependent subband separation, dispersion, and thermal occupation determine the diffusion coefficient.

\emph{Conclusion.---}
In summary, we have developed a microscopic and material-specific many-particle approach to describe exciton transport in the presence of an electronic Wigner crystal in atomically thin semiconductors. We show that in spite of a weak Wigner potential giving rise to only small energy shifts, it has a substantial impact on exciton transport. Considering the exemplary case of hBN-encapsulated MoSe$_2$ monolayers, we  predict the diffusion coefficients to be substantially decreased at low carrier densities. This behaviour is explained by a partial trapping of excitons in the periodic Wigner crystal potential that is reflected by a flattening of exciton subbands. We also show that exciton diffusion is both strongly density- and temperature-dependent when excitons interact with  Wigner crystal electrons. This opens up pathways for theoretical understanding of exciton transport in the presence of electronic correlations, predicting both strong and characteristic density-dependent effects, to be extended for a variety of scenarios introducing the physics of strongly correlated electrons to the field of exciton transport. 

\emph{Acknowledgments.---}
We thank Giuseppe Meneghini and Samuel Brem (Marburg University) as well as Raul Perea-Causin (Stockholm University) for fruitful discussions. This project has received funding from the  Deutsche Forschungsgemeinschaft via the regular project 542873285.

\end{document}


\renewcommand{\theequation}{S\arabic{equation}}
\renewcommand{\thesection}{\Roman{section}}
\preprint{APS/123-QED}

\title{Supplemental Material \---  }
\title{\textbf{Supplemental Material} \\ Impact of an electron Wigner crystal on exciton propagation}
\author{Daniel Erkensten$^{1,2}$, Alexey Chernikov$^{3}$ and Ermin Malic$^{1,2}$}
\affiliation{$^1$Department of Physics, Philipps-Universit{\"a}t Marburg, 35032 Marburg, Germany}
\affiliation{$^2$mar.quest|Marburg
Center for Quantum Materials and Sustainable Technologies,
35032 Marburg, Germany}
\affiliation{$^3$ Institute of Applied Physics and Würzburg-Dresden Cluster of Excellence ct.qmat,
TU Dresden, 01187 Dresden, Germany}
\maketitle 
\date{\today}

\section{Exciton-electron interaction}
We consider the local exciton-electron interaction $V_{x-e}(\mathbf{q})$ in the limit of low doping \cite{efimkin2021electron}
\begin{equation}
V_{x-e}(\mathbf{q})=\sum_{\mathbf{\mathbf{k}, \nu\neq 0 }}\frac{\Lambda^{0\nu}_{\mathbf{k}}\Lambda^{\nu 0}_{\mathbf{q}-\mathbf{k}}}{E_{0}^{X}-E^{X}_{\nu}}  \ , 
\label{localexc}
\end{equation}
with  the exciton energies $E^{X}_{\nu}$ and $\nu=(n,l)$, where $n$ and $l$ are the principal quantum number and the magnetic quantum number, respectively. Furthermore,  the scattering matrix element reads 
\begin{equation}
\Lambda^{\nu\nu'}_{\mathbf{k}}=U_{\mathbf{k}}\int \mathrm{d}^2\mathbf{r}(\varphi^{\nu}_{\mathbf{r}})^{*}\varphi^{\nu'}_{\mathbf{r}}2i\sin(\frac{\mathbf{k}\cdot\mathbf{r}}{2})\approx{i U_{\mathbf{k}} \mathbf{d}_{\nu\nu'}\cdot\mathbf{k}/e_0} \ , \hspace{1cm} (ka_x\ll 1)\ 
\end{equation}
introducing the Coulomb interaction $U_{\mathbf{k}}$, the transition dipole moment $\mathbf{d}_{\nu\nu'}=e_0\langle \nu |\mathbf{r}|\nu'\rangle$ and the real-space exciton wave function $\varphi^{\nu}_{\mathbf{r}}$. Considering small momenta compared to the inverse exciton Bohr radius $a^{-1}_x$, the exciton-electron interaction can be expressed as
\begin{equation}
V_{x-e}(\mathbf{q})=-\sum_{ \mathbf{k}, \nu\neq 0}\frac{U_{\mathbf{k}}U_{\mathbf{q}-\mathbf{k}}(\mathbf{d}_{0\nu} \cdot \mathbf{k})(\mathbf{d}^{*}_{0 \nu} \cdot (\mathbf{q}-\mathbf{k}))}{e_0^2(E_0^{X}-E^{X}_{\nu})}\ ,  \hspace{1cm} (ka_x\ll 1)
\end{equation}
which can be analytically evaluated in the long-wavelength limit ($q\rightarrow 0$):
\begin{equation}
\lim_{q\rightarrow 0}V_{x-e}(\mathbf{q})=-\frac{A}{4\pi e^2_0} \int_{0}^{1/a_x} \mathrm{d}k k^3 (U_k)^2 \alpha \ , \alpha=\sum_{\nu\neq 0 }\frac{|\mathbf{d}_{0\nu}|^2}{E_0^{X}-E^{X}_{\nu}} 
\end{equation}
with $A$ being the system area and $\alpha$ being the polarizability. 
The expression above can be evaluated for any form of the screened Coulomb interaction $U_k$. Note that we introduced a cut-off in the integral to be consistent with the approximation $ka_x\ll 1$. For the case of a pure 2D Coulomb interaction $U_k\sim 1/k$ we obtain 
\begin{equation}
\lim_{q\rightarrow 0}V_{x-e}(\mathbf{q})|_{2\mathrm{D}}=-\frac{\alpha e^2_0}{32\pi \epsilon_0^2 \epsilon_s^2 a_x^2A } \ . 
\label{2d}
\end{equation}
We may also evaluate the $q\rightarrow 0$ limit of the exciton-electron interaction assuming a screened Coulomb interaction of the Keldysh form \cite{keldysh1979coulomb, rytova1967screened} with $U_{k}=e_0^2/(2Ak\epsilon(k))$,  $\epsilon(k)=\epsilon_0 \epsilon_s(1+r_0 k)$. Here, $r_0$ is the material-specific screening length, $\epsilon_s$ the dielectric constant of the surrounding substrate. We find then
\begin{equation}
\lim_{q\rightarrow 0}V_{x-e}(\mathbf{q})|_{\mathrm{Keldysh}}=-\frac{\alpha e^2_0}{16\pi \epsilon_0^2 \epsilon_s^2 r_0^2A}\bigg[\mathrm{log}(1+\frac{r_0}{a_x})-\frac{r_0}{r_0+a_x}\bigg]\ .
\label{keldysh}
\end{equation}
In the limit $r_0\rightarrow 0$, Eq. \eqref{2d} and Eq. \eqref{keldysh} coincide as expected. The small momentum limit of the exciton-electron interaction enables us to find a quantitative estimate of the interaction strength. To get a material-specific value for the interaction, we evaluate Eq. \eqref{keldysh} and extract the polarizability from experiment \cite{cavalcante2018stark}. The exciton Bohr radius $a_x$ is microscopically computed from the Wannier equation and we find that $a_x\approx{1.5}$ nm in the considered case of hBN-encapsulated MoSe$_2$ monolayers, resulting in an exciton-electron coupling strength of $v_{x-e}\approx{-0.2}$ eV $\mathrm{nm}^2$ in the long wavelength limit (setting $A=1$). The exciton-electron interaction is set to this constant value throughout our work, corresponding to performing a contact potential approximation of the full exciton-electron interaction in real space (Fourier-transform of Eq. \eqref{localexc}). 
\label{geometry}

\section{Localization length of Wigner electrons}
We consider the electronic Hamilton operator 
\begin{equation}
\hat{H}_e=\sum_{i}[-\frac{\hbar^2 }{2m^*_e}\nabla^2_{\mathbf{r}_i}+\frac{1}{2}\sum_{j\neq i}U(\mathbf{r}_j-\mathbf{r}_i)] \ , 
\end{equation}
where $m^*_e$ is the effective carrier mass and $U(\mathbf{r}_i)$ is the Coulomb interaction. 
The total energy per lattice site $i$ is then given by $E/N=\langle \hat{H}_e\rangle/N=K+W$ (with $N$ being the total number of electrons) distinguishing between kinetic and potential contributions to the energy:
\begin{equation}
K=-\frac{\hbar^2}{2m^*_e}\int \mathrm{d}^2\mathbf{r}\psi^{*}_{i}(\mathbf{r})\nabla_{\mathbf{r}}^2\psi_{i}(\mathbf{r}) \ , W=\frac{1}{2}\int \mathrm{d}^2\mathbf{r}\mathrm{d}^2\mathbf{r}'U(\mathbf{r}-\mathbf{r}')|\psi_{i}(\mathbf{r})|^2\sum_{j\neq i}|\psi_{j}(\mathbf{r}')|^2 \ , 
\label{energycontributions}
\end{equation}
where we consider the Hartree contribution to the Coulomb interaction, which is the relevant contribution in the deep crystalline regime (away from Wigner crystal melting). Now, we perform the following Gaussian ansatz for the Wigner electron wave function:
\begin{equation}
\psi_{i}(\mathbf{r})=\frac{1}{\sqrt{2\pi\xi^2}}\mathrm{e}^{-|\mathbf{r}-\mathbf{R}_i|^2/(4\xi^2)} \ , 
\end{equation}
such that the Wigner crystal charge density reads $\rho(\mathbf{r})=\sum_{i}|\psi_{i}(\mathbf{r})|^2$. Now, we would like to find the localization length $\xi$. This is done by minimizing the total energy per site and treating $\xi$ as a variational parameter. Upon plugging the ansatz above into the kinetic energy in Eq. \eqref{energycontributions} we find $K=\frac{\hbar^2 }{4m^*_e\xi^2} $.
Without loss of generality, we can fix the site $i=0$ such that $\mathbf{R}_0=\mathbf{0}$, as the ansatz is valid for any site $i$. For the potential energy, we find 
\begin{align*}
W&=\frac{1}{2A(2\pi\xi^2)^2}\int \mathrm{d}^2\mathbf{r}\mathrm{d}^{2}\mathbf{r}'\sum_{j\neq 0 , \mathbf{q}}U_{\mathbf{q}}\mathrm{e}^{i\mathbf{q}\cdot(\mathbf{r}-\mathbf{r}'+\mathbf{R}_j)}\mathrm{e}^{-\frac{r^2}{2\xi^2}}\mathrm{e}^{-\frac{r'^2}{2\xi^2}}\\
&=\frac{n_e}{2(2\pi\xi^2)^2}\int \mathrm{d}^2\mathbf{r}\mathrm{d}^{2}\mathbf{r}'\sum_{\mathbf{g}}U_{\mathbf{g}}\mathrm{e}^{i\mathbf{g}\cdot(\mathbf{r}-\mathbf{r}')}\mathrm{e}^{-\frac{r^2}{2\xi^2}}\mathrm{e}^{-\frac{r'^2}{2\xi^2}}-\frac{1}{2A(2\pi\xi^2)^2}\int \mathrm{d}^2\mathbf{r}d^{2}\mathbf{r}'\sum_{\mathbf{q}}U_{\mathbf{q}}\mathrm{e}^{i\mathbf{q}\cdot(\mathbf{r}-\mathbf{r}')}\mathrm{e}^{-\frac{r^2}{2\xi^2}}\mathrm{e}^{-\frac{r'^2}{2\xi^2}}\\
&=\frac{n_e}{2}\sum_{\mathbf{g}\neq 0}U_{\mathbf{g}}\mathrm{e}^{-|\mathbf{g}|^2 \xi^2}-\frac{1}{4\pi}\int_{0}^{\infty}\mathrm{d}q q U_{\mathbf{q}}\mathrm{e}^{-\xi^2q^2} \ , 
\end{align*}
where we introduced the electron density $n_e=\frac{N}{A}$ and used $\sum_{j}\mathrm{e}^{i\mathbf{q}\cdot \mathbf{R_j}}=N\delta_{\mathbf{q}, \mathbf{g}}$ with $\mathbf{g}$ being reciprocal lattice vectors of the Wigner lattice. Hence, the total energy per site can be expressed as
\begin{equation}
E=\frac{\hbar^2}{4m^*_e\xi^2}+\frac{n_e}{2}\sum_{\mathbf{g}\neq 0}U_{\mathbf{g}}\mathrm{e}^{-|\mathbf{g}|^2 \xi^2}-\frac{1}{4\pi}\int_{0}^{\infty}\mathrm{d}q q U_{\mathbf{q}}\mathrm{e}^{-\xi^2q^2} \ .
\label{energymin}
\end{equation}
The energy  in Eq. \eqref{energymin} can be minimized with respect to the variational parameter $\xi$ for any lattice geometry and any form of the screened two-dimensional Coulomb interaction $U_ {\mathbf{q}}$. Note that in the first term, $\mathbf{g}=0$ is removed in order to cancel the contribution from the uniform positive background. In this work, we assume a Coulomb interaction of the Keldysh form \cite{keldysh1979coulomb, rytova1967screened}. However, we can gain additional information about the density dependence of the localization length by considering a Coulomb interaction of the simple purely two-dimensional form $U_{\mathbf{q}}\sim \frac{1}{\epsilon q}$ with $\epsilon$ being the effective dielectric constant of the material. By minimizing the energy in this case, i.e.,  setting $\frac{dE}{d\xi}=0$, we obtain the equation 
\begin{equation}
-\frac{\hbar^2}{2m^*_e}-\frac{\xi^4}{\sqrt{3}a_W^2 \epsilon}\sum_{\mathbf{g}\neq 0}|\mathbf{g}|\mathrm{e}^{-|\mathbf{g}|^2 \xi^2}+\frac{\xi}{\sqrt{\pi}\epsilon}=0 \ , 
\end{equation}
introducing the Wigner lattice constant $a_W$, such that $n_e^{-1}=A_W=\frac{\sqrt{3}}{2} a_W^2$ assuming a triangular lattice. Numerically, it is found that the third term is much smaller than the second one containing a sum over all reciprocal lattice vectors $\mathbf{g}\neq 0$ and therefore this term can be neglected. By noting that $|\mathbf{g}|\propto \frac{1}{a_W}$ and assuming that $\xi \ll a_W$, it then follows that $\xi \propto a_W^{3/4}\propto n_e^{-3/8}$. Furthermore, by introducing the localization ratio $\xi/a_W\propto n_e^{1/8}$, we thereby expect the Wigner electrons to be better localized at lower densities.

\label{geometry}
\section{Diagonalization of exciton-Wigner electron Hamiltonian}
The mean-field exciton-Wigner electron Hamiltonian reads in the exciton picture 
\begin{equation}
H=\sum_{\mathbf{Q}}E_{\mathbf{Q}}X^{\dagger}_{\mathbf{Q}}X_{\mathbf{Q}}+\sum_{\mathbf{q}, \mathbf{Q}}V_{x-e}(\mathbf{q})\rho_{e}(\mathbf{q})X^{\dagger}_{\mathbf{Q}+\mathbf{q}}X_{\mathbf{Q}} \ , 
\label{hamiltonian}
\end{equation}
where the first term contains the free parabolic center-of-mass exciton dispersion  $E_{\mathbf{Q}}=\frac{\hbar^2|\mathbf{Q}|^2}{2M}$ with $\mathbf{Q}$ being the center-of-mass momentum, and $M=m^{*}_e+m^{*}_h$ being the total exciton mass with $m^{*}_e$ and $m^{*}_h$ as electron and hole masses, respectively. The second term describes the mean-field interaction of excitons with Wigner crystal electrons, where the exciton-electron interaction $V_{x-e}(\mathbf{q})$ is weighted by the Wigner electron momentum charge density $\rho_e(\mathbf{q})$. The exciton-electron interaction can be approximated with a contact-like interaction in real space \cite{efimkin2021electron}, such that $V_{x-e}(\mathbf{q})=v_{x-e}$ (see Supplemental Section I). The Wigner electron charge density $\rho_e$ is assumed to have a Gaussian density profile in momentum and real space, i.e, 
$\tilde{\rho}_e(\mathbf{r})=\frac{1}{2\pi \xi^2}\sum_{n}\mathrm{e}^{-|\mathbf{r}-\mathbf{R}_n|^2/\xi^2}$ with $\mathbf{R}_n$ being real-space Wigner lattice vectors and $\xi$ the localization length (discussed in Supplemental Section II). It then follows that the momentum space charge density reads 
\begin{equation}
\rho_e(\mathbf{q})=\frac{1}{A_W}\sum_{\mathbf{g}}\mathrm{e}^{-\xi^2|\mathbf{g}|^2/2}\delta_{\mathbf{q},\mathbf{g}}\ ,
\end{equation}
where the Wigner unit cell area $A_W=\frac{\sqrt{3}}{2}a_W^2$ for a triangular lattice with $a_W$ being the Wigner lattice constant. Importantly, $\rho_{e}(\mathbf{q})=\sum_{\mathbf{g}}\bar{\rho}_e(\mathbf{g})\delta_{\mathbf{q}, \mathbf{g}}$, i.e., periodic with respect to the reciprocal lattice vectors $\mathbf{g}$ of the Wigner crystal. Given the periodicity of the exciton-Wigner electron potential, we diagonalize the Hamiltonian in Eq. \eqref{hamiltonian} by performing a zone-folding, i.e., taking the center-of-mass momentum $\mathbf{Q}\rightarrow \mathbf{Q}+\mathbf{g}$ and restricting the summation over $\mathbf{Q}$ to the (mini)-Brillouin zone of the Wigner crystal. Furthermore, by introducing new operators
 $Y^{(\dagger)}_{\eta, \mathbf{Q}}=\sum_{\mathbf{g}} C^{(*)}_{\eta, \mathbf{g}}(\mathbf{Q})X^{(\dagger)}_{\mathbf{Q}+\mathbf{g}}$, the Hamiltonian becomes diagonal such that  
\begin{equation}
\tilde{{H}}_{x,0}=\sum_{\eta, \mathbf{Q}}\tilde{E}^{\eta}_{\mathbf{Q}}Y^{\dagger}_{\eta, \mathbf{Q}}Y_{\eta, \mathbf{Q}} \ , 
\end{equation}
where $\tilde{E}^{\eta}_{\mathbf{Q}}$ is the renormalized exciton energy taking into account the exciton-Wigner electron potential. The renormalized exciton energies and the associated mixing coefficients $C_{\eta, \mathbf{g}}(\mathbf{Q})$ are obtained from solving the moiré eigenvalue problem
\begin{equation}
E_{\mathbf{Q}+\mathbf{g}}C_{\eta, \mathbf{g}}(\mathbf{Q})+\sum_{\mathbf{g}'}W(\mathbf{g}-\mathbf{g}')
C_{\eta, \mathbf{g}'}(\mathbf{Q})=\tilde{E}^{\eta}_{\mathbf{Q}}C_{\eta, \mathbf{g}}(\mathbf{Q}) \ ,
\label{eigenvalue}
\end{equation}
with $W(\mathbf{g})=V_{x-e}(\mathbf{g})\rho_e(\mathbf{g})$. The eigenvalue problem derived above is general and could be solved for any form of periodic interaction potential, $W$, and can be directly applied to excitons in a moiré potential induced by lattice mismatch or twisting in TMD-based heterostructures \cite{brem2020tunable}. 
Note that we, for simplicity and given the low temperatures considered in this work, consider only the lowest-lying bright (KK) excitons and their interaction with electrons in a Wigner crystal in MoSe$_2$ monolayers.

\section{Real-space Hamiltonian and localization of excitons} 
In real space, the mean-field exciton-electron Hamiltonian can be expressed as 
\begin{equation}
H_x=\sum_{i} H_{x,i}=\sum_{i}[-\frac{\hbar^2 \nabla^2_{\mathbf{r}_i}}{2M}+v_{x-e}\rho_i(\mathbf{r})] \ , 
\end{equation}
where $M$ is the total exciton mass (sum of electron and hole masses), $v_{x-e}$ is the exciton-electron interaction, here assumed to be a contact interaction, and 
\begin{equation}
\rho_i(\mathbf{r})=\frac{1}{2\pi \xi^2}\mathrm{e}^{-|\mathbf{r}-\mathbf{R}_i|^2/\xi^2} \ , 
\end{equation}
is the (Gaussian) real-space charge density of Wigner electrons. In the vicinity of the moiré lattice sites, i.e., $\mathbf{r}\approx{\mathbf{R}_i}$, we can approximate the mean-field interaction potential as a quantum harmonic oscillator potential and in particular (with $\mathbf{R}_0=0$), $\rho_0(\mathbf{r})\approx{\frac{1}{2\pi \xi^2}(1-\frac{r^2}{\xi^2})}$. In this case, it follows that the Wannier function associated with the lowest-lying exciton subband is Gaussian: 
\begin{equation}
\phi_{i}(\mathbf{r})=\frac{1}{\sqrt{\pi} a_x}\mathrm{exp}\bigg[-\frac{|\mathbf{r}-\mathbf{R}_i|^2}{2a_x^2}\bigg] \ , 
\end{equation}
with the spatial extent $a_x$. By demanding that $H_{x,0}\phi_{0}(\mathbf{r})=E \phi_{0}(\mathbf{r})$ for some constant energy $E$ we find that 
\begin{equation}
a_x=\bigg(\frac{\pi\hbar^2}{M|v_{x-e}|}\bigg)^{1/4}\xi \ , 
\end{equation}
i.e., the spatial extension of the exciton wave function is directly proportional to the localization length of the Wigner electrons. Hence, by combining this result with the scaling of the localization ratio with respect to Wigner electron density, $\xi/a_W\propto n_e^{1/8}$, we have shown that the excitonic wave function becomes delocalized as $n_e$ is increased. In other words, reducing the Wigner electron density is expected to result in a flattening of exciton bands. Notably, for excitons propagating in a moir{\'e} potential induced by a lattice-mismatch or a twist-angle in a TMD bilayer, it holds that $a_{x}/a_M\propto a_M^{-1/2}\propto n_M^{1/4}$ with $a_M$ being the moiré period and $n_M$ the moiré exciton density, i.e. a stronger density dependence is expected in such systems \cite{PhysRevLett.121.026402}. Given the weak exciton-electron potential ($|v_{x-e}|\approx 0.2$ eV nm$^2$), we find that the spatial extension of the exciton, $a_x$ is slightly smaller and of similar order of magnitude as the Wigner lattice constant $a_W$ for realistic Wigner electron densities ($n_e\sim 10^{11}$ $\mathrm{cm}^{-2}$).  

\section{Exciton-phonon scattering rates}
The periodic moir{\'e} potential that excitons feel and get trapped in due to the periodic Wigner lattice of electrons is analogous to the moir{\'e} potential induced by a lattice mismatch or by twisting in moir{\'e}  materials \cite{huang2022excitons}. In particular, the moir{\'e}  eigenvalue problem used to obtain the renormalized exciton energies and mixing coefficients (Eq. \eqref{eigenvalue}) is general and can be applied to any form of periodic interaction potential. Therefore, the recently established theory for exciton-phonon scattering in moir{\'e} structures can be directly used also in our work. The exciton-phonon scattering rate in a moiré system is derived within a second-order Born-Markov approximation and reads \cite{meneghini2024excitonic}
\begin{equation}
\Gamma^{\eta'\eta}_{\mathbf{Q}'\mathbf{Q}}=\frac{2\pi}{\hbar}\sum_{\pm, j, \mathbf{g}}|\tilde{G}^{\eta'\eta}_{j, \mathbf{Q}\mathbf{Q}'\mathbf{g}}|^2 (\frac{1}{2}\pm \frac{1}{2}+n_{j, \mathbf{Q}-\mathbf{Q}'+\mathbf{g}})\delta(\tilde{E}^{\eta}_{\mathbf{Q}}-\tilde{E}^{\eta'}_{\mathbf{Q}'}\pm \hbar\omega_{j, \mathbf{Q}-\mathbf{Q}'+\mathbf{g}}) \ ,
\end{equation}
where $\tilde{E}^{\eta}_{\mathbf{Q}}$ are the moir{\'e} exciton energies, $\eta$  the exciton subband index, $\mathbf{Q}$ the center-of-mass momentum, and $\mathbf{g}$ the reciprocal lattice vectors of the Wigner lattice. Furthermore, $n_{j, \mathbf{q}}$ is the Bose distribution for phonons, $\hbar\omega_{j, \mathbf{q}}$ are the phonon energies and the moir{\'e}  exciton-phonon matrix element reads \cite{meneghini2024excitonic, meneghini2025spatiotemporal} 
\begin{equation}
\tilde{G}^{\eta'\eta}_{j, \mathbf{Q}\mathbf{Q}'\mathbf{g}}=\sum_{\mathbf{g}', \tilde{\mathbf{g}}}G_{j, \mathbf{Q}-\mathbf{Q}'+\mathbf{g}}C^{*}_{\eta, \tilde{\mathbf{g}}}(\mathbf{Q})C_{\eta', \mathbf{g}'}(\mathbf{Q}')\delta_{\mathbf{g}, \mathbf{g}'-\tilde{\mathbf{g}}} \ , 
\end{equation}
with the exciton-phonon matrix element $G_{j, \mathbf{q}}$ and the moir{\'e} mixing coefficients $C_{\eta, \mathbf{g}}(\mathbf{Q})$. The mixing coefficients and the corresponding moir{\'e} eigenenergies are obtained from zone-folding and diagonalizing the Hamiltonian in Eq. (1) in the main manuscript with the details provided in Supplemental Section III.

The relaxation time $\tau^{\eta}_{\mathbf{Q}}$ that enters the exciton diffusion coefficient and the transport calculation is extracted directly from the phonon-driven out-scattering rate as 
\begin{equation}
(\tau^{\eta}_{\mathbf{Q}})^{-1}=\sum_{\eta', \mathbf{Q}'}\Gamma^{\eta'\eta}_{\mathbf{Q}'\mathbf{Q}} \ .
\end{equation}
The exciton-phonon coupling reads 
\begin{equation}
G_{j, \mathbf{q}}=g^{c}_{j, \mathbf{q}}F(\beta\mathbf{q})-g^{v}_{j, \mathbf{q}}F(-\alpha\mathbf{q}) \ , 
\end{equation}
with the electron-phonon matrix elements $g^{\lambda}_{j, \mathbf{q}}$ and the form factors $F(\mathbf{q})=\sum_{\mathbf{k}}\varphi^{*}_{\mathbf{k}+\mathbf{q}}\varphi_{\mathbf{k}}$ with $\varphi_{\mathbf{k}}$ being excitonic wave functions obtained from solving the Wannier equation \cite{brem2019intrinsic, kira2006many}. The mass ratio $\alpha=1-\beta=\frac{m^*_e}{m^*_e+m^*_h}$ is computed with the effective masses $m_e^*$ and $m_h^*$ from \emph{ab-initio} calculations \cite{kormanyos2015k}. We explicitly take into account intravalley exciton-phonon scattering involving longitudinal and acoustic phonon modes, $j=TA,LA$, relevant at the low temperatures considered in this manuscript \cite{selig2016excitonic}. Furthermore, the electron-phonon coupling matrix elements are treated within the deformation potential approximation \cite{PhysRevB.90.045422}:
\begin{equation}
g^{\lambda}_{j, \mathbf{q}}=\sqrt{\frac{\hbar}{2\rho \omega_{j, \mathbf{q}}}}D^{\lambda}_{j,\mathbf{q}} \ , 
\end{equation}
where $\rho$ is the surface mass density of the TMD monolayer, $A$ is the crystal area and $D^{\lambda}_{j,\mathbf{q}}$ is the deformation potential with $\lambda$ being the band index and $j$ the phonon mode. Note that the electron-phonon coupling for holes is related to that of valence band electrons as $D^{h}_{j, \mathbf{q}}=-D^{v}_{j, \mathbf{q}}$ and that $D^{h}_{j, \mathbf{q}}$ and $D^{e}_{j, \mathbf{q}}$ have the same sign, reflecting the non-polar nature of the interaction \cite{peelaers2012effects}. For long-wavelength acoustic phonons it holds that $D^{\lambda}_{j, \mathbf{q}}=\tilde{D}^{\lambda}_{j}\mathbf{q}$. Furthermore, the acoustic phonon frequencies read $\omega_{j, \mathbf{q}}=v_j |\mathbf{q}|$, where $v_j$ is the speed of sound. The constant deformation potential $\tilde{D}^{\lambda}_{j}$ and sound speed $v_j$ are obtained from DFPT calculations \cite{PhysRevB.90.045422}.
\section{Exciton diffusion in vicinity of Fermi sea of electrons}
\begin{figure}[t!]
\includegraphics[]{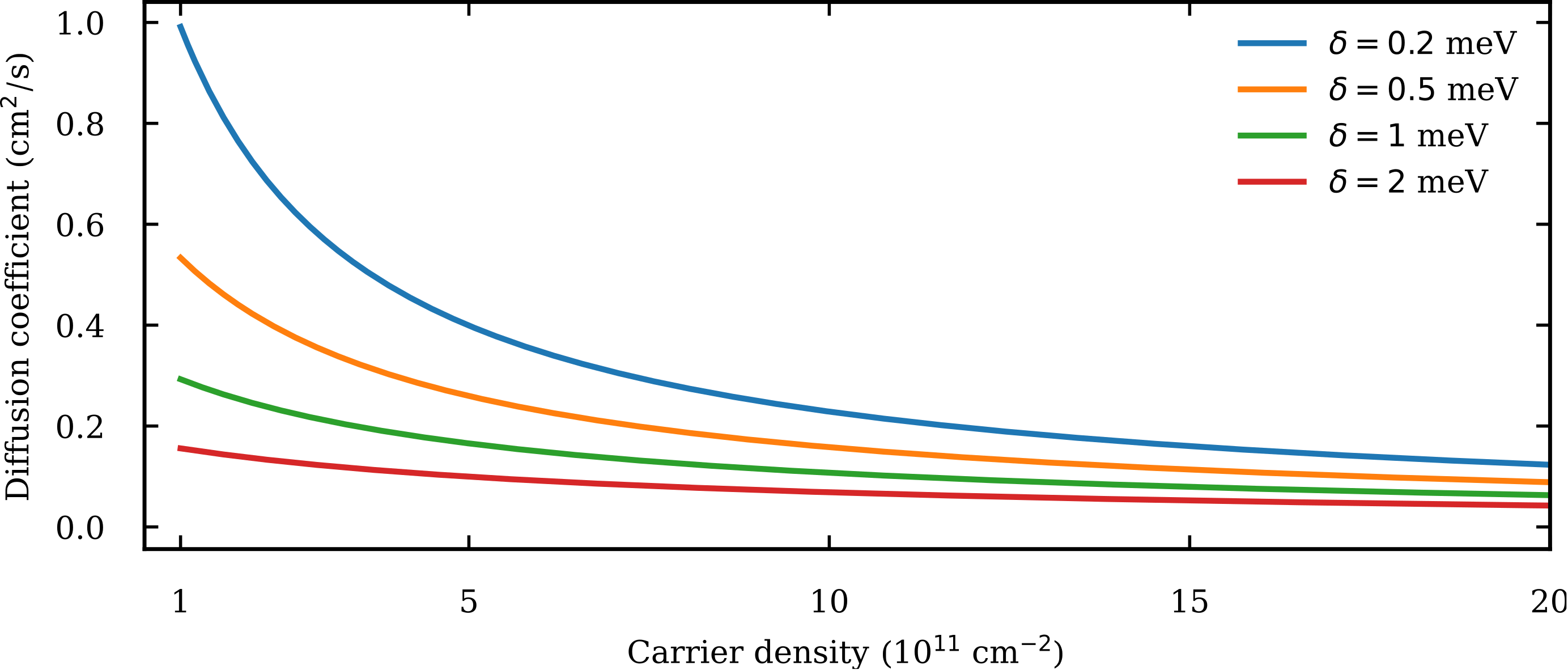}
\caption{Exciton diffusion in vicinity of Fermi sea of resident carriers as function of free carrier density for different values of the phonon-induced broadening $\delta$. }
\label{xdiff}
\end{figure}

The transport of excitons interacting with a Fermi sea of resident charge carriers can be studied using a Fermi-polaron approach as described in detail in Ref. \cite{wagner2023diffusion}. The linewidth broadening due to exciton-electron scattering is given by \cite{wagner2023diffusion} 
\begin{equation}
\label{gamma-x-e}
\Gamma_{x-e}(n_e)=\frac{n_e(m^*_e+M)\hbar^2\pi}{M m^*_e}\frac{\pi}{\mathrm{ln}^2(\delta/2E^b_t)+\frac{\pi^2}{4}} \ , 
\end{equation}
where $n_e$ is the free carrier density, $m^*_e$ is the resident electron mass, $M$ is the total exciton mass and $E^b_t$ is the trion binding energy. The latter can be extracted from experiments on n-doped hBN-encapsulated MoSe$_2$ monolayers yielding  $E^b_t=27$ meV \cite{zipfel2022electron}. The effective electron and hole masses are extracted from \emph{ab-initio} calculations \cite{kormanyos2015k} and are reported in Table I. Importantly, the exciton-electron scattering rate scales linearly with the carrier density or the Fermi energy. The parameter $\delta$ in Eq. \eqref{gamma-x-e} is unrelated to the exciton-electron interaction and describes exciton broadening due to, e.g., the coupling with phonons. It relaxes the energy and momentum conservation for exciton-electron scattering. The diffusion coefficient is then given by 
\begin{equation}
D=\frac{k_B T \hbar}{M[\delta+\Gamma_{x-e}(n_e)]} \ .
\label{diffusionequation}
\end{equation}
In Fig. \ref{xdiff}, we show the exciton diffusion coefficient of bright (KK) excitons in MoSe$_2$ using Eq. \eqref{diffusionequation} for different values of the  phonon-induced broadening $\delta$. We find a decrease of the diffusion coefficient with respect to density, reflecting the increased efficiency of the exciton-electron scattering at elevated densities.  In this work, we set  $\delta=0.2$ meV such that the diffusion coefficient at the lowest considered carrier density ($n_e=10^{11}$ $\mathrm{cm}^{-2}$) coincides with the predicted diffusion coefficient for a free exciton ($D\approx{1}$ cm$^2$/s). The opposite and qualitatively different density dependencies of exciton propagation in vicinity of an electron Wigner crystal and exciton transport in a Fermi sea hold also for a larger range of the broadening $\delta$.
\newpage 
\section{Input parameters}
The material-specific parameters extracted from \emph{ab initio} calculations are listed in Table \ref{parametertabell}. We provide the effective mass of electrons and holes $m^*$ (lowest-lying conduction band and highest-lying valence band at the K-point) in terms of the free electron mass $m_0$, the dielectric constant of hexagonal-boron nitride $\epsilon_s$ and the polarizability $\alpha$. Furthermore, we provide the relevant phonon parameters for the calculation of exciton-phonon scattering rates  including the speed of sound $v$ and electron and hole deformation potentials $\tilde{D}^{e}$ and $\tilde{D}^h$ of longitudinal and transverse acoustic phonons. \\
\begin{table}[h!]
    \centering
    \bgroup
\def\arraystretch{1.5}%
  \begin{tabular}{c|c|c}
\hline 
    Parameter  & Value & Ref.\\ \hline 
    Eff. Electron (Hole) mass $m^*$ [$m_0$] & 0.5 (0.6) & \cite{kormanyos2015k} \\ \hline  
    Dielectr. hBN $\epsilon_s$ & 4.5 & \cite{PhysRev.146.543} \\ \hline  
    Screening length $r_0$ [nm] & 1 & \cite{erkensten2024stability}\\ \hline 
      Polarizability $\alpha$ [eV nm$^2$/V$^2$] & 6.532 & \cite{cavalcante2018stark}\\ \hline 
       Speed of sound $v$ [cm/s] & 4.1$\cdot 10^{5}$ & \cite{PhysRevB.90.045422} \\ \hline 
       Electron (Hole) deformation potential $\tilde{D}$ [eV] & 3.4 (2.8) & \cite{PhysRevB.90.045422} \\ \hline 
\end{tabular}
\egroup
    \caption{Material-specific parameters used for exciton transport calculations in hBN-encapsulated MoSe$_2$ monolayers. The phonon parameters are given for LA/TA phonons and are assumed to be the same for the longitudinal and transverse modes.}
    \label{parametertabell}
\end{table}